\documentclass{llncs}
\usepackage[usenames,dvipsnames]{color}
\usepackage{xspace}
\definecolor{mygray}{gray}{0.30}
\usepackage{listings}
\usepackage{amsmath}
\usepackage{amssymb}
\usepackage{graphicx}
\lstdefinelanguage{mypseudo}{
morekeywords={while,if,otherwise, interface, extends, where, return, for, true, false},
otherkeywords={::,_},
sensitive=true,
morecomment=[l]{--},
commentstyle={\color{mygray}\itshape}
}
\newcommand{\pred}[1]{\ensuremath{\textrm{#1}}}

\newcommand{\eqprop}{\textbf{\textsf{allequal}}}

\begin{document}
\title{Integrating Datalog and Constraint Solving} 
\author{Benoit~Desouter \and Tom~Schrijvers}
\institute{Ghent University, Belgium \\ \email{\{Benoit.Desouter,Tom.Schrijvers\}@UGent.be}} 
\maketitle
\begin{abstract}
LP is a common formalism for the field of databases and CSP, both at the theoretical level and the implementation level in the form of Datalog and CLP. In the past, close correspondences have been made between both fields at the theoretical level. Yet correspondence at the implementation level has been much less explored. In this article we work towards relating them at the implementation level. Concretely, we show how to derive the efficient Leapfrog Triejoin execution algorithm of Datalog from a generic CP execution scheme.
\end{abstract}

\section{Introduction}
Constraint programming (CP) is a well-known paradigm in which relations
between variables describe the properties of a solution to the problem we wish
to solve \cite{programmingWithConstraints}. The strategy how to actually compute these solutions is left to the system.
Databases have been an interesting research topic since the 1960's.
Constraint programming and databases span two separate domains, each with their own insights and techniques. They are not immediately similar. However database theory and CP, in particular CLP, actually have much in common~\cite{vardi}, at least at the theoretical level. A common language for them is first-order logic, which does not involve any computational aspects.

Cross-fertilization between the two could give us more expressive systems and better results.
Hence we look at logic programming as a common computational language: 
Datalog is a query language for deductive databases used in a variety of
applications, such as retail planning, modelling, \ldots~\cite{datalogNeverAsk}.

This paper aims to show that Datalog and CP are
also compatible at the implementation level. We do so by showing how a standard
CP implementation scheme, as formulated by Schulte and Stuckey~\cite{efficientProp}, can be
specialized to obtain a recently documented Datalog execution algorithm called
\emph{Leapfrog Triejoin}~\cite{leapfrogTriejoin}.
Leapfrog Triejoin has a good theoretical complexity and is simple to implement.

This integration opens up the possibility for further
cross-fertilization between actual CP and Datalog systems. In particular, we
aim to integrate CP propagation techniques in the Datalog join algorithm for
query optimization.

%

\section{The Abstract Constraint Solving Scheme} \label{sec:abstractScheme}

\newcommand{\search}[1]{\textbf{\textsf{search}}(#1)}
\newcommand{\isolv}[1]{\textbf{\textsf{isolv}}(#1)}
\newcommand{\searchnoarg}{\textbf{\textsf{search}}}
\newcommand{\isolvnoarg}{\textbf{\textsf{isolv}}}
Our starting point is the abstract algorithm for efficient CP propagator
engines as formulated by Schulte and Stuckey~\cite{efficientProp}, listed in
Figure~\ref{fig:solver}.

The inputs to the algorithm are a set of old (constraint) propagators $F_o$, a
set of new propagators $F_n$ and the domain $D$ of the constraint variables.
Initially the set of old propagators is empty and a toplevel calls takes the form
$\search{\emptyset,F_n,D}$. These inputs are derived from a declarative problem
specification that relates a finite set of constraint variables $\mathcal{V}$ by
means of a number of constraints $C$: 
\begin{itemize}
\item
All of the variables have an initial set of admissible values, which are
captured in $D$.  This domain $D$ is a mapping from $\mathcal{V}$ to the
admissible values; we denote the set associated with a variable $x$ as $D(x)$.
For example, if the admissible values for $x$ are the integers from one to four, we write $D(x) = \{ 1, 2, 3, 4 \}$.
\item
The constraints are captured in a set of constraint propagators $F$ (typically
one or several per constraint). Such a propagator $f$ is a monotonically
decreasing function on domains that removes values that do not feature in
any possible solution to the constraint.

Define, in addition to $x$ and its domain defined above, a variable $y$ with $D(y) = \{ 3, 4, 5 \} $. A constraint propagator for $x = y$ can then eliminate the values $\{ 1, 2 \} $ from $D(x)$ and 5 from $D(y)$.
\end{itemize}

Given these inputs it is the algorithm's job to figure out whether the
constraint problem has a solution.
To do so, it alternates between two phases: \emph{constraint propagation}
and \emph{nondeterministic choice}. 

Constraint propagation computes a fixpoint of the constraint propagators;
it is captured in the function $\isolv{F_o,F_n,D}$ which we will explain
in more detail below. This may yield one of three possible outcomes:
\begin{enumerate}
\item One of the variables has no more admissible values. Then there is no
      solution.
\item All of the variables have exactly one admissible value. A solution
      has been found.
\item At least one of the variables has two or more admissible values.
\end{enumerate}
The first two cases terminate the algorithm. The last case leads to
nondeterministic choice. The current search space $C \wedge D$ is partitioned
using a set of constraints $\{c_1,\ldots,c_n\}$. Typical approaches include the use of two constraints that each split the domain of a     
certain variable in half, or constraints that either remove or assign a value. A large number of strategies for choosing a split variable or a value exists. One may pick the variable with the largest domain, the smallest domain, \ldots, the smallest value or a random one, etc. In all of those cases each of the subspaces
is explored recursively in a depth-first order. The $i$th recursive subcall
gets $F_o \cup F_n$ as old propagators\footnote{We know that $D$ is a fixpoint of them.} and the new propagators of $c_i$ as the new
ones.

\begin{figure}[t]
\fbox{
\begin{minipage}{\textwidth}
\begin{center}
\begin{tabbing}
xxx \= xx \= xx \= xx \= xxxxxxxxxxxxxxxxxxxxxxxxxxxxxxxxxxxxxx \= xxx \kill
\search{$F_o$, $F_n$, $D$} \\
  \> $D$ := \isolv{$F_0$,$F_n$,$D$} \> \> \> \> \textit{\% propagation} \\
  \> \textbf{if} ($D$ is a false domain) \\ 
  \> \> \textbf{return} \\
  \> \textbf{if} ($\exists x \in \mathcal{V}.\left\vert D(x) \right\vert > 1 $) \\
  \> \> choose $ \{c_{1},\ldots, c_{m} \} $ where  $C \wedge D \models c_1 \vee \cdots \vee c_m $ \>\>  \>  \textit{\% search strategy} \\
  \> \> \textbf{for} $ i \in [1 .. m] $ \\
  \> \> \> \search{$F_o \cup F_n$, prop($c_i$), $D$} \\
  \> \textbf{else} \\
  \> \> \textbf{yield} $D$ \\
\end{tabbing}
\end{center}
\end{minipage}
}
\caption{General constraint solver\label{fig:solver}}
\end{figure}

\paragraph{Incremental Constraint Propagation} \label{subsec:propphase}

Figure~\ref{fig:isolv} shows an incremental algorithm for constraint propagation.
It takes a set of old propagators $F_o$, new propagators $F_n$ and a constraint domain
as inputs, and returns a reduced domain as output. The invariant is that for every old propagator
$f_o \in F_o$, $D$ is a fixpoint (i.e., $D = f_o(D)$). The propagators that may still
reduce $D$ are in $F_n$; they are used to initialize a worklist $Q$.

Then the algorithm repeatedly takes a propagator $f$ from $Q$ and uses it to
obtain a possibly reduced domain $D'$. Then an auxiliary function (not given)
$\textsf{new}(f,F,D,D')$ determines what new propagators from $F$ to add to the
worklist; these should be propagators for which $D'$ may not be a fixpoint. A
valid but highly inefficient implementation of \textsf{new} just returns $F$,
but typical implementations try to be more clever and return a much smaller set
of propagators.

When the worklist is empty, the algorithm returns $D$ which is now a fixpoint
of all propagators $F$.
 
\begin{figure}[t]
\begin{center}
\fbox{
\begin{minipage}{\textwidth}
\begin{tabbing}
xxx \= xx \= xxx \= xxx \= xxxxxxxxxxxxxxxxxx \= xxx \kill
\isolv{$F_o$, $F_n$, $D$} \\
  \> $F$ := $F_o \cup F_n$; $Q$ := $F_n$; \\
  \> \textbf{while} ($ Q \neq \emptyset $) \\
  \> \> $f$ := first($Q$) \\
  \> \> $Q$ := $ Q - \{f\} $; $ D' := f(D) $ \\
  \> \> \textbf{if} ($ D' \neq D $) \\
  \> \> \> $ Q := \textsf{new}(f,F,D,D')$ \\
  \> \> $ D := D' $ \\
  \> \textbf{return} $D$ 
\end{tabbing}
\end{minipage}
}
\end{center}
\caption{Incremental constraint propagation\label{fig:isolv}}
\end{figure}

\subsection{Instantiation}

In practice the generic scheme is instantiated to fill in unspecified details
(like how the partition is obtained) and refined to obtain better efficiency.
For instance, when $D$ is reduced by a propagator, typically not all variables are affected.
The \textsf{new} function would only return those propagators
that depend on the affected variables. Moreover, efficient pointer-based datastructures
would be used to quickly identify the relevant propagators.

In the rest of the paper we will apply various such instantiations and refinements.
Yet our goal is not to obtain a concrete CP system. Instead, we have as target
the Datalog Leapfrog-Triejoin execution algorithm.


\section{The Datalog Instance}
Datalog execution uses rules to derive new facts from known facts. A rule has the form
$$ \textrm{h} \leftarrow \textrm{b}_{1}, \ldots, \textrm{b}_{n} $$
where $\textrm{h}, \textrm{b}_{1}, \ldots, \textrm{b}_{n}$ are atomic formulas. An atomic formula has the form
$$\textrm{p}(X_1, \ldots, X_n)$$
 where p is a predicate with arity $n$ and the $X_i$ are variables.
Every predicate refers to a table of facts of the form
$ \textrm{p}(\textrm{c}_1, \ldots, \textrm{c}_n) $ 
with the $\textrm{c}_i$ constants.
If the body is instantiated by known facts, then the head yields a (possibly) new fact.

The most performance-critical part of the instantiation is the \emph{join} which finds facts that share a common argument.
Suppose we have the following facts:
$$ \textrm{p(a,b)}, \textrm{p(c,d)}, \textrm{p(e,f)} $$
$$ \textrm{q(a,1)}, \textrm{q(c,2)}, \textrm{q(g,3)} $$
Then the join $ \pred{p}(X,Y), \pred{q}(X,Z) $ gives us the following results: $ \{X \mapsto a , Y \mapsto b, Z \mapsto 1\} $ and $ \{X \mapsto c , Y \mapsto d, Z \mapsto 2\} $.

As is clear from the example, a join between three unary predicates looks like
$$ \pred{p}(X), \pred{q}(X), \pred{r}(X)$$
 We can rewrite the rule body to the following form
$$ \pred{p}(X), \pred{q}(Y), \pred{r}(Z), X = Y = Z$$ 
that makes the equalities explicit. Now the following analogy with constraint satisfaction problems becomes more obvious:
\begin{itemize}
\item The rule variables correspond to constraint variables.
\item The predicates denote the domains of the variables.
\item The equalities are constraints on the variables.
\end{itemize}
Note that Datalog only uses one kind of constraints: the \emph{global equality constraint}. A generic propagator for this constraint is shown in~Figure~\ref{fig:genericEquality}, that only performs propagation on the lower bound.

It introduces a variable mapping $M$ from $ \{0, \ldots, n - 1\} $ to $\mathcal{V}$. The mapping is essentially an array of pointers to the elements of $D$. We sort $M$ by increasing lower bound of the variables in $D$. Then, the variable $x$ pointed to by the last position in $M$ has the largest lower bound $l_{\textrm{max}}$. For each of the other variables, we eliminate values smaller than $l_{\textrm{max}}$. If this operation leads to a larger lower bound, we start the entire process again. If, in contrast, the lower bound of all variables is equal to $l_{\textrm{max}}$, we have found a fixpoint from which we can derive a solution. Thus the algorithm maintains the invariant that the lower bounds of the variables pointed to by the array elements at indices $i \ldots (i + n) \bmod n$ are a sorted series.

As an example consider three variables $X, Y$ and $Z$ with respective domains $D(X) = \{ 1, 2, 3, 4, 9, 10, 11 \}, D(Y) = \{ 3 , 4, 7, 10 \}$ and $D(Z) = \{ 1 , 4, 7, 10, 11 \}$. Sorting $M$ by increasing lower bound then gives us $ \{ 0 \mapsto X , 1 \mapsto Z, 2 \mapsto Y \}$. In Table~\ref{tab:example}, we illustrate the process, underlining the domain with the largest lower bound. The absence of a value means that there are no changes with respect to the previous line in the table.

 Initially $l_{\textrm{max}} = 3$. In the first iteration, we increase the lower bound of $X$ to $3$ and $l_{\textrm{max}}$ does not change. In the next iteration, the lower bound of $Z$ is increased to $4$. The maximum lower bound $l_{\textrm{max}}$ is updated accordingly. In the next two iterations the lower bounds of $Y$ and $X$ are again increased to end up at $4$. We now have found a solution.

\begin{figure}[t]
\begin{center}
\fbox{
\begin{minipage}{\textwidth}
\begin{tabbing}
xxx \= xx \= xxx \= xxx \= xxxxxxxxxxxxxxxxxx \= xxx \kill
\eqprop($D$) \\
\> make a variable mapping $M = \{ 0 \mapsto x_{1}, \ldots, n - 1 \mapsto x_{n} \}$ to $D$ \\
\> sort $M$ by increasing lower bound in $D$ \\
\> $l_{\textrm{max}}$ := lowerBound($D[M[n - 1]]$) \\
\> $i$ := 0 \\ 
\> \textbf{while} (lowerBound($D[M[i]]$) $\neq$ $l_{\textrm{max}}$) \\
\> \> $D[M[i]]$.raiseLowerBound($l_{\textrm{max}})$ \\
\> \> $ l_{\textrm{max}} $ := lowerBound($D[M[i]]$) \\
\> \> $i$ := $(i + 1) \bmod n$ \\
\> \textbf{return} $D$
\end{tabbing}
\end{minipage}
}
\end{center}
\caption{Generic equality propagator\label{fig:genericEquality}}
\end{figure}

\begin{table}
\begin{equation*}
\begin{array}{ccc}
\boldsymbol{X} & \boldsymbol{Z} & \boldsymbol{Y} \\
\{ 1, 2, 3, 4, 9, 10, 11 \}             & \{ 1 , 4, 7, 10, 11 \}             & \underline{\{ 3 , 4, 7, 10 \}} \\
\underline{\{ 3, 4, 9, 10, 11 \}}       &                                    &   \\
                                        & \underline{\{ 4, 7, 10, 11 \}}     &   \\
                                        &                                    & \underline{ \{ 4, 7, 10 \}} \\
\underline{\{4, 9, 10, 11 \}}           &                                    &   \\

\end{array}
\end{equation*}
\caption{Example operation of the propagator for global equality\label{tab:example}}
\end{table}

\subsection{Unary Datalog}
When we restrict ourselves to unary Datalog, we only solve CP problems with a single equality propagator at a time.
For rules with multiple variables like
$$ \pred{p}(X), \pred{q}(X), \pred{r}(Y), \pred{s}(Y)$$
we calculate one join at a time. The final result is then the Cartesian product of the solutions for $X$ and $Y$.

In this situation, \isolvnoarg\ is trivial to implement as a single invocation of the propagator. This is valid because the propagator is idempotent:
$$\eqprop(\eqprop(D)) = \eqprop(D)$$ 

\section{Making a choice}
The abstract constraint solving scheme from Section~\ref{sec:abstractScheme} does not specify how 
to add extra constraints $c_{i}$ to $D$ when propagation alone does not yield a solution. Recall that the set of constraints $c_{i}$ added in turn must partition the search space. A well-known technique, the indomain-min strategy, is to select a variable $x$ and either assign or remove its lower bound $lb$: $c_{1} \equiv (x = lb), c_{2} \equiv (x > lb)$. 

This technique is particularly attractive here because the propagator has already made sure that all variables have the same lower bound. Thus assigning the lower bound of one variable with $c_{1}$ requires no work. In particular it requires no further propagation by the \eqprop\ propagator, so we immediately have a solution that we can yield.

In the other branch, we increment the domain. That means we eliminate the lower bound from a random variable. We then continue as before. We do not need an additional propagator for $c_{2}$: by eliminating the lower bound $lb$, the constraint $c_{2}$ is satisfied right away. 

\begin{figure}
\begin{minipage}{0.5\textwidth}
\begin{center}
\includegraphics[width=0.8\textwidth]{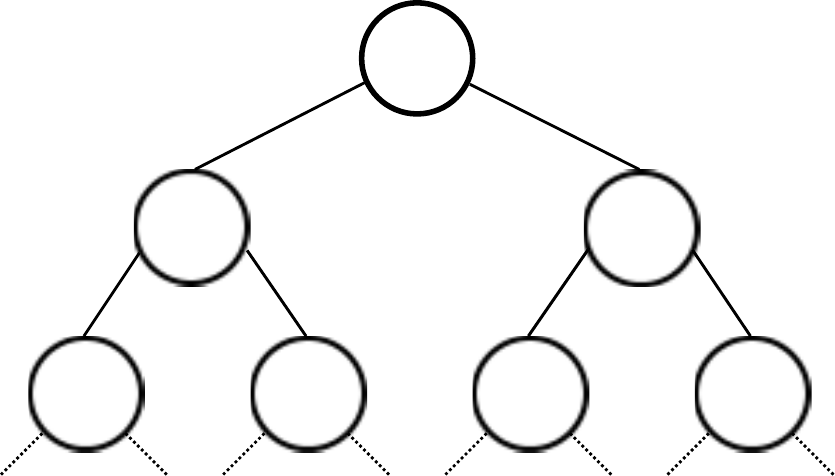}
\end{center}
\end{minipage}
\begin{minipage}{0.5\textwidth}
\begin{center}
\includegraphics[width=0.8\textwidth]{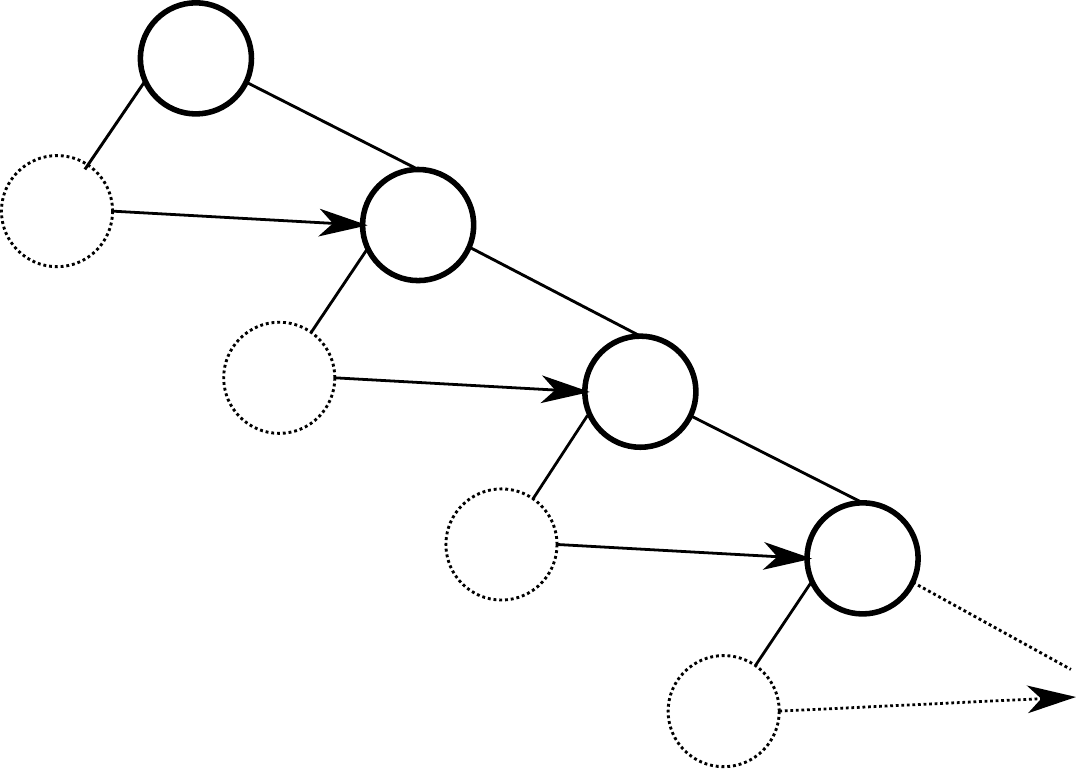}
\end{center}
\end{minipage}
\caption{General search tree (left) vs. tree searched by our tail recursive algorithm (right).\label{fig:trees}}
\end{figure}

In Figure~\ref{fig:searchspecial2} we show the impact of this refinement on the specialized constraint solver. Note that the algorithm is now tail recursive and thus can easily be turned into a \textbf{while} loop that runs in constant stack space. Contrast this with conventional CP systems that need a stack to perform depth first search. We illustrate the difference in Figure~\ref{fig:trees}. On the left is a general search tree; on the right the tree searched in our code.
The dashed nodes represent the solution found after the call to \eqprop. From this node, we can move on to the rest of the tree by following the dashed arrow. It corresponds to the incDomain operation.

As an example of the approach, again consider the three variables $X$, $Y$ and $Z$ with the same domains as above. The first solution is $X = Y = Z = 4$. After yielding this solution, the domains are $X = \{ 4, 9, 10, 11 \}, Y = \{ 4, 7, 10 \}$ and $Z = \{ 4, 7, 10, 11\}$. We now increase the lower bound of variable $X$. During the next iteration of the while loop, \textbf{allequal} is applied again to find the solution $X = Y = Z = 10$. Once again incrementing the lower bound of $X$ leaves us with an empty domain and the while loop terminates.

\begin{figure}[t]
\fbox{
\begin{minipage}{\textwidth}
\begin{center}
\begin{tabbing}
xxx \= xx \= xx \= xx \= xxxxxxxxxxxxxxxxxxxxxxxxxxxxxxxxxxxxxx \= xxx \kill
\search{$D$} \\
  \> $D$ := \eqprop($D$) \> \> \> \> \textit{\% propagation} \\
  \> \textbf{if} ($D$ is a false domain) \\ 
  \> \> \textbf{return} \\
  \> \textbf{yield} $D$ \> \> \> \> \textit{\% lower bounds equal} \\
  \> $D$ := incDomain($D$,$x$) \\ 
  \> \search{$D$} \\
\end{tabbing}
{\Large $\Downarrow$ }
\begin{tabbing}
xxx \= xx \= xx \= xx \= xxxxxxxxxxxxxxxxxxxxxxxxxxxxxxxxxxxxxx \= xxx \kill
\search{$D$} \\
\> \textbf{while} (\textit{true}) \\
\>  \> $D$ := \eqprop($D$) \\

\> \> \textbf{if} ($D$ is a false domain) \\ 
\> \> \> \textbf{return} \\
\> \> \textbf{yield} $D$ \\
\> \> $D$ := incDomain($D$,$x$) \\  
\end{tabbing}
\end{center}
\end{minipage}
}
\caption{Effect of the indomain-minimum value selection on the constraint solver\label{fig:searchspecial2}}
\end{figure}

\section{Leapfrog Triejoin}
When we inline the code for the \eqprop\ constraint from Figure~\ref{fig:genericEquality} within the constraint solver with indomain-min value selection, it is clear we can introduce one more optimization. Indeed, we do not need to resort the variable domains on every invocation of the propagator. This is because we know the variable modified by the incDomain operation must be the one that has the new largest lower bound. All other variables have not been modified since we found a solution. To avoid having to change the position $p_{\textrm{min}}$ where the solution was detected, it is most convenient to increase the variable at position $p_{\textrm{min}} - 1 \bmod n$ in $M$. In that way, there is absolutely no work involved in maintaining the ordering. The result can be seen in Figure~\ref{fig:fullintegrated}. The algorithm is now exactly the same as Leapfrog Triejoin.

We begin by sorting the array of pointers $M$ to variables $x$ by increasing lower bound. As before, we keep the maximum lower bound in $l_{\textrm{max}}$. The variable $x_{\textrm{min}}$ having the smallest lower bound can be found at position $p_{\textrm{min}}$ in $M$.
As before, we know we have a solution if $l_{\textrm{min}}$ is equal to $l_{\textrm{max}}$. The inner while loop either stops because this is the case, or because there is a variable $x$ with an empty domain. In the latter case, all solutions have been found.
In the former case, we yield the solution and increment the domain. This is done in such a way that we can immediately start the inner while loop again.

\begin{figure}[t]
\begin{center}
\fbox{
\begin{minipage}{\textwidth}
\begin{tabbing}
xxx \= xx \= xxx \= xxx \= xxxxxxxxxxxxxxxxxx \= xxx \kill
\search{$D$} \\
\> make a variable mapping $M$ to $D$ \\
\> sort $M$ by increasing lower bound in $D$ \\
\> $p_{\textrm{min}}$ := 0 \\
\> $l_{\textrm{max}}$ := lowerBound($D[M[n-1]]$) \\
\> \textbf{while} ($D$ is not a false domain) \\
\> \> $foundSolution$ := \textit{false} \\
\> \> \textbf{while}($\lnot$ ($foundSolution$ $\vee$ $D$ is a false domain)) \\
\> \> \> $x_{\textrm{min}}$ := $D[M[p_{\textrm{min}}]]$ \\
\> \> \> $l_{\textrm{min}}$ := lowerBound($x_{\textrm{min}}$) \\
\> \> \> \textbf{if} ($l_{\textrm{min}} = l_{\textrm{max}}$) \\
\> \> \> \> $foundSolution$ := \textit{true} \\
\> \> \> \textbf{else} \\
\> \> \> \> $l_{\textrm{max}}$ := $x_{\textrm{min}}$.raiseLowerBound($l_{\textrm{max}}$) \\
\> \> \> \> $p_{\textrm{min}}$ := $(p_{\textrm{min}} + 1) \bmod n $ \\
\> \> \textbf{if} ($foundSolution$) \\
\> \> \> \textbf{yield} $D$ \\
\> \> \> $D$ := incDomain($D$,$D[M[p_{\textrm{min}} - 1 \bmod n]]$) \\
\end{tabbing}
\end{minipage}
}
\end{center}
\caption{Leapfrog Triejoin algorithm\label{fig:fullintegrated}}
\end{figure}

\section{Full Datalog Implementation}

\paragraph{Iterator implementation}
We have started from an abstract domain representation $D$. In CP it is typically represented as a union of intervals $ \bigcup [lb_{i},ub_{i}]$ where $lb_{i+1} > ub_{i} + 1$. We only use a restricted set of operations in the algorithm of Figure~\ref{fig:fullintegrated}:
\begin{itemize}
\item Access to the lower bound from the domain of a variable $x$.
\item Removing that lower bound from the domain.
\item Removing all values smaller than a certain value from the domain.
\end{itemize}
In a Datalog context, tables are normally stored as trees. But as described in Veldhuizen's work~\cite{leapfrogTriejoin}, it is perfectly possible to implement the necessary operations on top of trees. The resulting concept is called an iterator.

\paragraph{N-Ary predicates}
In addition to the operations needed in the unary case, an iterator offers three additional operations for working with general predicates: \textbf{open} and \textbf{up} are used to move in the tree-based representation of a relation. From a higher level, we can describe this as moving between the variables in a predicate. The function \textbf{depth} then indicates which variable we are currently manipulating.

The basic approach for non-unary predicates is to use one Leapfrog Triejoin per set of variables that must be equal. For example, if 
$\textrm{p}(X,Y)$ and $\textrm{q}(Z,Q)$ are two binary predicates and we join on $X = Z, Y = Q$, we first calculate a solution for $X = Z$. Given this configuration, all solutions for $Y = Q$ are looked for. Then we look for the next solution where $X = Y$ and repeat the entire process.

\paragraph{Datalog System}
A fully functional Datalog system has the ability to store the new facts derived by the program rules. This can be achieved by collecting the answers and storing them in trees. Recursive rules can be handled with a semi-naive algorithm. Both capabilities do not   influence the core algorithm described in this paper.

\section{Related Work}
Much work has been done in coupling logic programming languages to relational databases. The oldest method, relational access, lets Prolog access only one table at a time and combines data from multiple tables using depth-first search. It is clear that this method is very inefficient, since it does not exploit any of the optimizations from the DBMS. Maier \textit{et al.}~\cite{prologRelationalNed2002,prologRelationalNed2003} have stressed the importance of achieving this coupling efficiently. A more recent approach thus translates Prolog database access predicates into appropriate SQL queries~\cite{draxlerPrologToSql}. Although arguably more efficient, the integration may have varying degrees of transparency. Queries are generally isolated from the rest of the Prolog program. Therefore, they may not use all information available in the Prolog program to restrict the number of records accessed even more. Furthermore, not all queries expressible in Prolog can be translated to SQL.

Compared to our work, these integration techniques are rather loosely coupled. Back in 1986, Brodie and Jarke stated that tightly integrating logic programming techniques with those of DBMSs will yield a more capable system. They estimated this requires no more work than extending either with some facilities of the other~\cite{brodie86}. Unfortunately, to date, no full integration of Prolog and relational databases has gained a significant degree of acceptance~\cite{prologRelationalNed2002}. Datalog, on the other hand, has been successfully used as a more integrated approach.

\section{Conclusions and Future Work}
\label{sec:concl}
The integration of Datalog and constraint programming offers many interesting perspectives in join optimization.
In this article, we have only described the core ideas behind this integration.

In future, we will first and foremost generalize the approach for non-unary predicates.
Intuitively leapfrog triejoin for non-unary predicates corresponds to nested searches. This only allows for propagation between the arguments in one direction. Techniques that allow for more propagation between the arguments definitely deserve our attention. 

Furthermore, we will also investigate both impact and advantages of adding propagators for additional constraints. A well-known example is the $X < Y$ constraint. Consider this constraint and the domains $X = \{ 1, 8 \}$ and $Y = \{ 2, 3, 5, 6, 9  \}$. It is clear that after finding all solutions where $X = 1$, one can at once discard the values $ \{ 2, 3, 5, 6 \} $ from the domain of $Y$.

When the less-than constraint is used together with a join, as in
$$\pred{p}(X), \pred{q}(Y), \pred{r}(Y), X < Y$$
we would now first do the join on $\pred{q}(Y)$ and $\pred{r}(Y)$ and then filter out the values where $X < Y$. It is clear we can improve here with more propagation.  
Finally, many standard constraint programming optimizations can still be added to the system. 

\section*{Acknowledgments}
We would like to thank LogicBlox, Inc. for their support and for giving us the ability to investigate the integration of Datalog and constraint programming in their system.



\bibliography{bib}

\end{document}